\documentclass[11pt]{article}

\let\ssection=\section
\renewcommand{\section}{\setcounter{equation}{0}\ssection}
\usepackage{bm}% bold math
\usepackage{amsmath,dsfont}

\usepackage{graphics,amssymb,amsmath,amsthm,amscd,array}
\usepackage{graphicx,subfigure}

\def\parag{\hfil\break} %%%%% paragraph
\def\kikezd{\parag\underbar}

\def\IR{{\mathds{R}}} %%%%% Reals

\def\p{{\partial}}

\def\vp{{\vec{p}}}
\def\vr{{\vec{r}}}
\def\vP{{\vec{P}}}
\def\vR{{\vec{R}}}
\def\vk{{\vec{k}}}
\def\vJ{{\vec{J}}}
\def\vL{{\vec{L}}}
\def\vK{{\vec{K}}}
\def\vM{{\vec{M}}}
\def\vv{{\vec{v}}}
\def\vV{{\vec{V}}}
\def\vU{{\vec{U}}}
\def\vA{{\vec{A}}}
\def\vB{{\vec{B}}}

\def\cL{{\cal{L}}}
\def\vnabla{{\vec{\nabla}}}

\def\vpi{{\vec{\pi}}}
\def\vN{{\vec{N}}}

\def\smallover#1/#2{\hbox{$\textstyle\frac{#1}{#2}$}}

\def\dAlembert{\vcenter {
    \hbox {\vrule height8pt width0.4pt depth0.0pt
           \vrule height8pt width7.2pt depth-7.6pt
           \vrule height8pt width0.4pt depth0.0pt
           \kern-8pt
           \vrule height0.4pt width8pt depth0.0pt
          \,}}}%--- Francisco's box
\def\IR{{\mathds{R}}} %%%%% Reals

\def\IS{{\mathds{S}}} %%%%%
\def\ri{\mathrm{i}} %%%%%
\def\bv{\mathbf{v}}

\title{{\bf Dynamical symmetry of the Kaluza-Klein monopole}\footnote{
Talk given at the '88 Schloss Hofen Meeting on {\it Symmetries in Science III}.
Gruber B and Iachello F (eds). Plenum~: New York.
p. 399-417 (1989).}
}

\author{L. Feh\'er\footnote{
Bolyai Institute, University of Szeged.
Present address: Research Institute for Particle and Nuclear Physics, Budapest, Hungary.
e-mail: lfeher-at-rmki.kfki.hu}
\quad
and
\quad
P. A. Horv\'athy\footnote{Dipartimento di Fisica, Universit\`a di Napoli, Italy.
Present address: Laboratoire de Math\'ematiques et de Physique Th\'eorique, Universit\'e de Tours, France.
e-mail: horvathy-at-lmpt.univ-tours.fr}
}
\begin{document}

\maketitle

\begin{abstract}
The Kepler-type dynamical symmetries of the Kaluza-Klein monopole are reviewed. At the classical level, the conservation of the angular momentum  and of a Runge-Lenz vector imply that the trajectories are conic sections.
The ${\rm o}(4)$ algebra allows us to calculate the bound-state spectrum, and
the ${\rm o}(3,1)$ algebra yields the scattering matrix.
The symmetry algebra extends to ${\rm o}(4,2)$.
\end{abstract}

\noindent
Arxiv:0902.4600

%%%%%%%%%%%%%%%%%%%%%%%%%%%%%%%%%%%%%%%%%%%%%%%%%%
\section{SUMMARY OF KALUZA-KLEIN THEORY~\cite{1,2}}
%%%%%%%%%%%%%%%%%%%%%%%%%%%%%%%%%%%%%%%%%%%%%%%%%%

One of the oldest and most enduring ideas regarding the unification of gravitation
and gauge theory is Kaluza's five dimensional unified theory. Kaluza's hypothesis was
that the world has four spatial dimensions, but one of the dimensions has curled up
to form a circle so small as to be unobservable. He showed that ordinary general
relativity in five dimensions, assuming such a cylindrical ground state, contained
a local U(1) gauge symmetry arising from the isometry of the hidden fifth dimension.
The extra components of the metric tensor constitute the gauge fields of this
 symmetry and could be identified with the electromagnetic vector potential.

To be more specific, consider general relativity on a five dimensional
 space-time with the Einstein-Hilbert action
\begin{equation}
S=-\frac{1}{16\pi G_K}\int d^5x\sqrt{-g_5}\,R_5
\label{(1.1)}
\end{equation}
where $R_5$ is the five-dimensional curvature scalar of the metric
$g_{AB}$, $g_5={\rm det}(g_{AB})$, and
$G_K$ is the five-dimensional coupling constant. Our conventions are: upper case Latin letters $A, B, C$ denote five-dimensional indices $0, 1, 2, 3, 5$; lower case Greek indices $\mu, \nu \dots$ run over four dimensions,
$0, 1, 2, 3$, whereas lower case Latin indices run over four-dimensional spatial values $1, 2, 3, 5$. The signature of $g_{AB}$ is
$(-,+,+,+,+)$ and the Riemann tensor is
$$
\begin{array}{l}
R^K_{\ LMN}=\p_M\Gamma^K_{\ LN}-\p_N\Gamma^K_{\ LM}+
\Gamma^K_{\ JM}\Gamma^J_{\ LN}-\Gamma^K_{\ JN}\Gamma^J_{\ LM}
\\[8pt]
R_{LM}=R^K_{\ LKM},\qquad
R_5=R^L_{\ L},
\end{array}
$$
and, except where indicated, $h/2\pi=c=1$.
In the absence of other fields the equations of motion are of course
$R_{AB}=0$.

The basic assumption of Kaluza and Klein was that the correct vacuum is the space $M^4\times \IS^1_R$, the product of four dimensional Minkowski space with a circle of radius $R$. The radius of the circle in the fifth dimension is undetermined by the classical equations of motion, since any circle is flat. If $R$ is sufficiently small then all low-energy experiments will simply average over the fifth dimension. In fact the components of the metric, $g_{AB}(x^\mu, x^5)$, can be expanded in Fourier series,
\begin{equation}
g_{AB}(x^\mu,x^5)=\sum_ng_{AB}^{(n)}(x^\mu)\exp\left[\frac{\ri nx^5}{R}\right],
\label{(1.2)}
\end{equation}
and all modes with $n\neq0$ will have energies greater than $hc/R$. Thus the effective low-energy theory can be deduced by considering the metric
$g_{AB}$ to be independent of $x^5$. Under these assumptions the theory is invariant under general coordinate transformations that are independent of
$x^5$. In addition to ordinary four dimensional coordinate transformations
$x^\mu\to x^\mu(x^\nu)$, we have a U(1) local gauge transformation
$x^5\to x^5+\Lambda(x^\mu)$, under which $g_{\mu5}$ transforms as a vector gauge field,
\begin{equation}
g_{\mu5}(x)\to g_{\mu5}(x)+\p_\mu\Lambda.
\label{(1.3)}
\end{equation}
Therefore the low-energy theory should be a theory of four dimensional
gravity plus a $U(1)$ gauge theory, i.e. electromagnetism, with the massless modes of
$g_{\mu\nu}$, $g_{\mu5}$ corresponding to the graviton (photon). The low-energy theory is also
invariant under scale transformations in the fifth dimension,
\begin{equation}
        x^5 \to \lambda x^5, \qquad
        g_{55}\to \lambda^{-2}g_{55},\qquad
        g_{\mu5} \to  \lambda^{-1} g_{\mu5}.
\label{(1.4)}
\end{equation}
This global scale invariance is spontaneously broken by the Kaluza-Klein vacuum (since R is fixed) thus giving rise to a Goldstone boson, the dilaton.

To exhibit the low-energy theory we write the metric as follows
\begin{equation}\begin{array}{lll}
        g_{AB}&=&\left(\begin{array}{cc}
        g_{\mu\nu} + A_\mu A_\nu &A_\mu V
        \\[10pt]
        A_\nu V &V
        \end{array} \right),
        \\[20pt]
        g_5&=&{\rm det}g_{AB}={\rm det}(g_{\mu\nu}) V=g_4 V,
        \\[10pt]
        ds^2&=&V(dx^5+A_\mu dx^\mu)^2+g_{\mu\nu}dx^\mu dx^\nu.
        \end{array}
\label{(1.5)}
\end{equation}
The five-dimensional curvature scalar can be expressed in terms of the four dimensional curvature,
$R_4$, the field strength,
$F_{\mu\nu}=\partial_{\mu}A_{\nu}-\partial_{\nu}A_{\mu}$, and the scalar field,
 $V$,
\begin{equation}
        R_5=R_4 + \smallover{1}/{4}VF_{\mu\nu}F^{\mu\nu}-\frac{2}{\sqrt{V}}\,\dAlembert\sqrt{V}\ .
\label{(1.6)}
\end{equation}
Thus the effective low energy theory is described by the four dimensional action
\begin{equation}
        S  =  -\frac{1}{16\pi G}\int d^4x\sqrt{-g_4}\,V^{1/2}\left(R_4 + \smallover{1}/{4}VF_{\mu\nu}F^{\mu\nu}\right),
\label{(1.7)}
\end{equation}
where we have dropped the terms in $R_5$ involving $V$, since,  when multiplied by $\sqrt{g_5}$,
these yield a total derivative, and
\begin{equation}
        G=\frac{G_K}{2\pi R}
\label{(1.8)}
\end{equation}
is Newton's constant, determined by $\sqrt{hG}/\sqrt{2\pi c^3}
\approx 1.6\times10^{-33}$ cm.

    This theory is recognizable as a variant of the Brans-Dicke theory \cite{3} of gravity, with $V^{1/2}$ identified as a Brans--Dicke massless scalar field, coupled to electromagnetism. $V$ indeed sets the local scale of the gravitational coupling. In the vacuum $V=1$. Also in the Brans--Dicke theory the coupling of $V$
to matter is somewhat arbitrary, here it is totally fixed by five-dimensional covariance.
    The radius $R$ is determined by the electric charge. To see this consider a complex field $\Phi$ with action
\begin{equation}
        S_{\Phi}=\int d^5x \sqrt{-g_5}\,(\partial_A\Phi)(\partial^A\Phi^\dagger).\label{(1.9)}
\end{equation}
The Fourier component of $\Phi$ with non-trivial $x^5$ dependence,
$\Phi^{(n)}(x^\mu) \exp[\ri nx^5/R]$, will behave as a particle of charge
$e =\sqrt{16\pi G}n/R$ and mass $n/R$, since for $\Phi(x^\mu, x^5)=\Phi^{(n)}(x^\mu)\exp[\ri nx^5/R]$
\begin{equation}
        \partial_A\Phi\partial^A\Phi^\dagger =\left|\left(\partial_\mu+i\frac{n}{R}A_\mu
        \right)\Phi^{(n)}\right|^2+\frac{n^2}{VR^2}\big|\Phi^{(n)}\big|^2.
\label{(1.10)}
\end{equation}
(Note that the properly normalized gauge field is $(16\pi  G)^{-1/2}A_\mu)$. Thus
\begin{equation}
             \alpha=e^2/2hc=2h G/\pi c^3R^2,
             \qquad
             R=\sqrt{\frac{2hG}{\pi \alpha c^3}}
             \approx  3.7 \times 10^{-32} {\rm cm}.
\label{(1.11)}
\end{equation}
    Consider a classical point test particle of unit mass. It has action
\begin{equation}
        S=\int d\tau\sqrt{g_{AB}\frac{dx^A}{d\tau}\frac{dx^B}{d\tau}}\
\label{(1.12)}
\end{equation}
and the motion is given by a five-dimensional geodesic,
\begin{equation}
\frac{d^2x^A}{d\tau^2}+\Gamma^A_{\ BC}\,\frac{dx^B}{d\tau\; }\frac{dx^C}{d\tau\; }=0.
\label{(1.12half)}
\end{equation}
Now, since the space-time possesses a Killing vector, namely
\begin{equation}
 K^A \frac{\ \partial}{\partial x^A} =\frac{\ \partial}{\partial x^5},
\label{(1.13)}
\end{equation}
it is guaranteed that $K_Adx^A/d\tau$ is a constant of the motion. Indeed a first integral of eqn. (\ref{(1.12half)}) is
\begin{equation}
    K_A\frac{dx^A}{d\tau}=V\frac{dx^5}{d\tau}+A_\mu\frac{dx^{\mu}}{d\tau}
    =q.
\label{(1.14)}
\end{equation}
The remaining equations of motion then take the form
\begin{equation}
        \frac{d^2x^\mu}{d\tau^2}+\Gamma^\mu_{\alpha\beta}\frac{dx^\alpha}{d\tau}
        \frac{dx^\beta}{d\tau}=qF^\mu_{\ \nu}\frac{dx^\nu}{d\tau}+
        q^2\,\frac{\partial^\mu V}{2V^2}.
        \label{(1.15)}
\end{equation}
Here $\Gamma^\mu_{\alpha\beta}$ is the four dimensional connection constructed from
$g_{\mu\nu}$. We recognize on the right the Lorentz force if
$q$ is identified with the charge of the particle, plus an interaction with the scalar field $V$.

%%%%%%%%%%%%%%%%%%%%%%%%%%%%%%%%%%%%%%%%%%%%%%%%%%%%%%%%%
\section{THE MONOPOLE OF GROSS, PERRY AND SORKIN~\cite{2}}
%%%%%%%%%%%%%%%%%%%%%%%%%%%%%%%%%%%%%%%%%%%%%%%%%%%%%%%%%

    Now we are searching for solitons in the five-dimensional Kaluza - Klein theory sketched above. By a soliton we mean a non-singular solution to the classsical field equations which represent spatially localized lumps that are topologically stable. Such solutions are expected to have a large mass (of the order
$M/g$, where $M$ is the mass scale of the theory and $g$ a dimensionless coupling constant). They are the starting points for the semi-classical construction of quantum mechanical particle states.

Our goal is to construct solutions of the five dimensional field equations that approach the vacuum solution: $V = 1,\ A_\mu = 0,\ g_{\mu\nu}=\eta_{\mu\nu}$ at spatial infinity. It is natural to consider static metrics with
$\partial/\partial t$ as Killing vector. It is also natural to look for solutions with $g_{0A} =\delta_{0A}$. In this case the space-time is totally flat in the `time' direction and the field equations are simply
\begin{equation}
        R_{ij}=R_{5i}=R_{55}=0,
\label{(2.1)}
\end{equation}
namely the four-dimensional, wholly space-like, manifold at each fixed $t$ has a vanishing Ricci tensor. These are simply the equations of four dimensional euclidean gravity, where we can think of $x^5$ as representing euclidean, periodic time. Our task is greatly simplified by the fact that the equations of four-dimensional euclidean gravity have been extensively studied. For example, the Kaluza-Klein monopole of Gross and Perry, and of Sorkin \cite{2}, which is the object of our considerations here, is obtained by imbedding the Taub-NUT gravitational instanton into five dimensional Kaluza-Klein theory. Its line element is expressed as
\begin{equation}
ds^2 =-dt^2 +\left(1+\frac{4m}{r}\right)\Big(
dr^2+r^2(d\theta^2 +\sin^2\!\theta\, d\phi^2)\Big)+
\displaystyle\frac{(d\psi + 4m\cos\theta\, d\phi)^2}{1+ \displaystyle\frac{4m}{r}}\ ,
\label{(2.2)}
\end{equation}
where $r > 0$, and the angles $\theta, \phi, \psi,\
(0\leq\theta\leq\pi, 0\leq\phi\leq2\pi)$ parametrize
$\IS^3 \approx SU(2)$. The apparent singularity at the origin is unphysical \cite{4}
if
$\psi$ is periodic with period $16\pi m$. Since we want our solution to approach the vacuum for large $r$, we must identify $16\pi m$ with $2\pi R$. Thus
\begin{equation}
        m=\frac{R}{8}=\frac{\sqrt{\pi G}}{2e}\ .
\label{(2.3)}
\end{equation}

The gauge field, $A_\mu$, is clearly that of a monopole,
$A_\phi=4m\cos\theta, \vB = \vnabla \times \vA = 4m\, \vr/r^3$, and has a string singularity along the whole $z$ axis.
As usual, this singularity is an artifact if and only in the period of $x^5$ is equal to $16\pi m$.
      The magnetic charge of our monopole is thus fixed by the radius of the Kaluza-Klein circle. If we scale the magnetic field so as to have the proper normalization,
$\vB\to (16\pi G)^{-1/2}\vB$, we find that the magnetic charge is
\begin{equation}
    g=\frac{4m}{\sqrt{16\pi G}}=\frac{R}{2\sqrt{16\pi G}}=
    \frac{1}{2e}.
\label{(2.4)}
\end{equation}
Thus, as expected, our monopole has one unit of Dirac charge.

    The mass of a static, asymptotically flat spacetime can be defined. In our case it is
\begin{equation}
M=-\frac{2\pi R}{16\pi G_K}\,\int d^3x\vnabla^2\big(\frac{1}{V}\big)=\frac{m}{G}\ .
\label{(2.5)}
\end{equation}
Since m is fixed by the radius of the vacuum circle, which in turn is fixed by $e$ and $G$, the soliton mass is determined to be
\begin{equation}
    M^2=\frac{m_P^2}{16\alpha}\, ,
\label{(2.6)}
\end{equation}
where $m_P=\sqrt{hc/2\pi G} \approx 2.17 \cdot 10^{-5}$ g is the Planck mass. As it is costumary for solitons, the monopole mass is
$1/e$ times heavier than the mass scale of the theory.

    Remarkably, the Kaluza-Klein monopole has re-emerged recently \cite{5} as the asymptotic limit of the curved manifold, whose geodesics describe the scattering of self-dual monopoles.

%%%%%%%%%%%%%%%%%%%%%%%%%%%%%
\section{CLASSICAL DYNAMICS}
%%%%%%%%%%%%%%%%%%%%%%%%%%%%%

    Let us consider the geodesic motion of a particle in the Kaluza -- Klein monopole field, with Lagrangian
\begin{equation}
        \cL = \frac{1}{2}g_{\mu\nu}\dot{x}^\mu\dot{x}^\nu
        =\frac{1}{2}\left((1+\frac{4m}{r})\dot{\vr}\,{\strut}^2+
        \displaystyle\frac{(\dot{\psi}+4m\cos\theta\dot{\phi})^2}
        {1+\displaystyle\frac{4m}{r}}\right).
\label{(3.1)}
\end{equation}
To the two cyclic variables $\psi$ and $t$ are associated the conserved quantities
\begin{eqnarray}
    q&=&\frac{\dot{\psi}+4m\cos\theta\dot{\phi}}{1+\displaystyle\frac{4m}{r}},
\label{(3.2a)}
\\[8pt]
    E&=&\frac{1}{2}(1+\frac{4m}{r})\left(\,\dot{\vr}\,{\strut}^2+q^2\right),
\label{(3.2b)}
\end{eqnarray}
interpreted as the electric charge and the energy, respectively. It is convenient to introduce the mechanical 3-momentum
\begin{equation}
        \vp=\big(1+\frac{4m}{r}\big)\dot{\vr}.
        \label{(3.3)}
\end{equation}
The 5-dimensional geodesics are the solutions of the Euler-Lagrange equations associated to (\ref{(3.1)}).
The projection into 3-space of the motion is hence governed by the equation,
\begin{equation}
    \frac{d\vp}{dt}=-4mq\frac{\dot{\vr}\times\vr}{r^3}
    +2mq^2\frac{\vr}{r^3}-2m\,\dot{\vr}\,{\strut}^2
    \frac{\vr}{r^3}.
\label{(3.4)}
\end{equation}
This complicated equation contains, in addition to the Dirac-monopole plus  Coulomb terms, also a velocity-square dependent term, typical for motion in curved space.

Due to the manifest spherical symmetry, the monopole angular momentum,
\begin{equation}
        \vJ=\vr\times\vp+ 4mq \frac{\vr}{r},
\label{(3.5)}
\end{equation}
is conserved. The presence of the velocity-square dependent force changes,
when compared to the pure Dirac + Coulomb case, the situation dramatically.
Energy conservation implies  that
\begin{equation}
    (\dot{r})^2=\frac{(2E-q^2)r^2-(E-q^2)r-J^2}{(r+4m)^2}.
\label{(3.6)}
\end{equation}
For positive $m$ the particle cannot reach the center. Indeed, eqn. (\ref{(3.2b)}) shows that, for  $m>0$ the energy is at least $q^2/2$, and hence
\begin{equation}
    r \geq r_0=\frac{4mq^2}{2E-q^2}>0.
\label{(3.7)}
\end{equation}

For $m > 0$ there are no bound motions.

    For $m < 0$ instead, the energy can be smaller than $q^2/2$.
In such a case the coefficient of $r^2$ is negative and the system does admit bound motions (see below).

\begin{figure}
\begin{center}
\includegraphics[scale=.5]{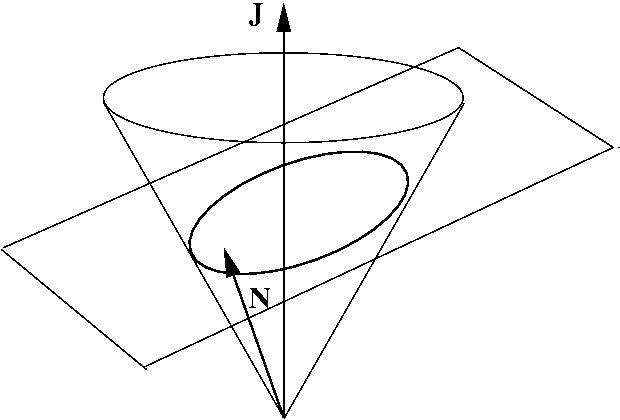}
\end{center}
\caption{\it The particle moves on a cone, whose axis is the conserved angular momentum,
${\vJ}$. The position belongs to a plane
 perpendicular to
${\vN}=q{\vK}+(E-q^2){\vJ}$. The trajectory is, therefore, a conic section.}
\label{Fig1}
\end{figure}

    Despite the complicated form of the equations of motion, the classical motions are surprisingly simple. The clue is the observation \cite{6} that, in addition to the angular momentum,
$\vJ$, there is also a conserved `Runge-Lenz' vector, namely
\begin{equation}
        \vK  =  \vp\times\vJ-4m(E-q^2)\frac{\vr}{r}.
\label{(3.8)}
\end{equation}
This is verified by an explicit calculation, using the equations of motion.
    These conserved quantities allow for a complete description of the motion
  \cite{6}. Indeed, eqns. (\ref{(3.5)}) and (\ref{(3.8)}) allow us to prove that
\begin{eqnarray}
        \vJ\cdot\frac{\vr}{r}&=& 4mq,
\label{(3.9a)}
\\[8pt]
\left(\vK+\frac{(E-q^2)}{q}\,\vJ\right)\cdot\vr&=&J^2-(4mq)^2.
\label{(3.9b)}
\end{eqnarray}
The first of these equations implies that, as it is usual in monopole
interactions, the particle moves on a cone with axis $\vJ$ and opening angle
 $\cos\alpha=\vert 4 mq\vert/J$. The second implies in turn that the motions lie in the plane perpendicular to the vector
$\vN=q\vK+(E-q^2)\vJ$.
They are therefore conic sections, see Fig. \ref{Fig1}.

\noindent
The form of the trajectory depends on $\beta$, the plane's inclination, being smaller or larger then the
complement of the cone's opening angle, $\pi/2-\alpha$. Now
$\cos\beta=\vN\cdot\vJ/NJ$. Using the relations
\begin{eqnarray}
    \vK\cdot\vJ&=&-(4m)^2q(E-q^2),
\label{(3.10a)}
\\[8pt]
    \vK^2&=&(2E-q^2)(\vJ^2-(4mq)^2)+4m^2(E-q^2)^2,
\label{(3.10b)}
\end{eqnarray}
a simple calculation yields that for
$$
\left.\begin{array}{lll}
    E  &<  &q^2/2
    \\[6pt]
    E &=&q^2/2
    \\[6pt]
    E  &>  &q^2/2
    \end{array} \right\}
    \qquad\hbox{the trajectories are}\qquad
    \left\{\begin{array}{lll}
    \hbox{ellipses}
    \\[6pt]
    \hbox{parabolae}
    \\[6pt]
    \hbox{hyperbolae}
    \end{array}\right.
$$

    Let us now study the unbound motions. Using the two conserved quantities $\vJ$ and $\vK$
the classical scattering can also be described. Let $\omega$ be the scattering angle, $\gamma$ the
 twist, and let $\vec{\bv}$ denote the velocity at large distances. Then
$E=(\vec{\bv}^2+q^2)/2$. The modulus of the orbital angular momentum,
$L=|\vL|$, is $L= y\vert\vec{\bv}\vert$, where $y$ is the impact parameter.
From the conservation of the component of $\vJ$ which is parallel to the particle's plane, we get,
\begin{equation}
        \tan\frac{\pi-\theta}{2} = - y\vert\vec{\bv}\vert\frac{\sin\gamma}{4qm}.
\label{(3.11)}
\end{equation}
From the conservation of the parallel component of $\vK$, we get
in turn
\begin{equation}
    \tan\frac{\pi-\theta}{2}=-y\vert\vec{\bv}\vert\frac{\cos\gamma}{4m(E-q^2)} .
    \label{(3.12)}
\end{equation}
From these equations we deduce
\begin{equation}
    \tan\frac{\gamma}{2}=\frac{q}{\vert\vec{\bv}\vert}
    \qquad\hbox{and}\qquad
    \tan\frac{\theta}{2}=-\frac{2m}{y}\left(1+\left(\frac{q}{\vert\vec{\bv}\vert}\right)^2\right).
\label{(3.13)}
\end{equation}
The classical cross-section is, therefore,
\begin{equation}
\frac{d\sigma}{d\omega}=y\frac{|dy/d\theta|}{\sin\theta}=
\frac{m^2}{\sin^4\theta/2}\left(1+\left(\frac{q}{\vert\vec{\bv}\vert}\right)^2\right).
\label{(3.14)}
\end{equation}

To clarify the structure of the conserved quantities, it is  convenient to switch to the Hamiltonian formalism. The momenta which are canonically conjugate to the coordinates are \hfill\break
$\pi_i=(1+4m/r)\dot{x}_i-qA_i,\ i=1,2,3$,  and $\pi_4=q$, where
$A_r=A_\theta=0,\  A_\phi=-4m\cos\theta$. The Poisson brackets are
$$
\big\{x^i,\pi_j\big\}=\delta^i_{\ j},
\qquad
\big\{\psi,\pi_4\big\}=1.
$$
The mechanical momentum $\vp=\vpi+q\vA=(1+4m/r)\dot{\vr}$ satisfies therefore the Poisson bracket relations
\begin{equation}
     \big\{p_i, p_j\big\}=-(4qm)\epsilon_{ijk}\frac{x^k}{r^3},
     \qquad
     \big\{x^i, p_j\big\}=\delta^i_{\ j},
     \qquad
     \big\{\psi,p_j\big\}=A_j.
\label{(3.15)}
\end{equation}
The Hamiltonian is
\begin{equation}
    H=\frac{1}{2}\left(\frac{\vp\,{\strut}^2}{1+\displaystyle\frac{4m}{r}}+(1+\frac{4m}{r})q^2\right).
\label{(3.16)}
\end{equation}
The Poisson brackets of the angular momentum and the Runge-Lenz vector are
\begin{equation}
     \big\{J_i,J_k\big\}=\epsilon_{ikn}J_n,
     \qquad
     \big\{J_i, K_k\big\}=\epsilon_{ikn}K_n,
     \qquad
     \big\{K_i, K_k\big\}=(q^2-2H)\epsilon_{ikn}J_n .
\label{(3.17)}
\end{equation}
For each fixed value $E$ set
\begin{equation}
\vM\quad=\;\left.\begin{array}{r}
(q^2-2E)^{-1/2}\vK
\\[6pt]
\vK
\\[6pt]
(2E-q^2)^{-1/2}\vK
\end{array}\right\}
\qquad\hbox{for}\qquad
\left\{\begin{array}{c}
E<q^2/2
 \\[6pt]
E=q^2/2
\\[6pt]
E>q^2/2
\end{array}\right.\ .
\label{(3.18)}
\end{equation}
For energies lower then $q^2/2$ the angular momentum, $\vJ$, and the rescaled Runge-Lenz
vector, $\vM$, form an ${\rm o}(4)$ algebra \cite{6}. For $E>q^2/2$ the
dynamical symmetry generated by $\vJ$ and $\vM$ is rather ${\rm o}(3,1)$.
In the parabolic case, $E=q^2/2$, the algebra is ${\rm o}(3)\oplus_s\IR^3$.

%%%%%%%%%%%%%%%%%%%%%
\section{QUANTIZATION}
%%%%%%%%%%%%%%%%%%%%%

    First, we find the quantum operators by DeWitt's rules \cite{7}.
    
$\hat{x}_i$ is multiplication by $x_i$, and the canonical momentum operator is
$$
    \hat{\pi}_i=-\ri\big(\partial_i+\smallover1/4(\partial_i\ln g)\big).
$$
$\hat{x}^i$ and $\hat{\pi}_i$ are hermitian with respect to $L^2(\sqrt{g}d^4x)$,
and satisfy the basic commutation relations
$$
[\hat{x}^i,\hat{x}^j]=
[\hat{\pi}_i,\hat{\pi}_j]=0,
\qquad
[\hat{x}^i,\hat{\pi}_j]=\ri\delta^i_{\ j}\,.
$$
 It is, however, more useful to introduce the modified (with respect to the Taub-NUT volume element non-self-adjoint!)
 operator
\begin{equation}
    \hat{p}_k=-\ri\partial_k+qA_k,
\label{(4.1)}
\end{equation}
with commutation relations
\begin{equation}
    [\hat{p}_i,\hat{p}_j]=-\ri(4mq)\epsilon_{ijk}\frac{x^k}{r^3},
        \qquad
    [\hat{x}^i,\hat{p}_j]=\ri\delta^i_{\ j},
        \qquad
    [\hat{\psi},\hat{p}_j]=\ri A_j\,.
    \label{(4.2)}
\end{equation}
The charge operator $\hat{q} = -\ri\partial_\psi$ has eigenvalues 
\begin{equation}
q = s/4m,\ s = 0, \pm 1/2, \pm 1,\dots
\end{equation}
 Our 4-metric is Ricci flat, so the quantum Hamiltonian is \cite{7}
\begin{equation}
    \widehat{H}=\frac{1}{2}g^{-1/4} \hat{\pi}_\mu\sqrt{g}\,g^{\mu\nu}\hat{\pi}_\nu g^{-1/4}
    =-\frac{1}{2\sqrt{g}}\,\partial_\mu\big(\sqrt{g}\,\partial^\mu\big)=-
    \frac{\bigtriangleup}{2},
\label{(4.3)}
\end{equation}
the covariant Laplacian on curved 4-space. Then a lengthy calculation yields the three dimensional expression
\begin{equation}
    \widehat{H}=\frac{1}{2}\left(\frac{1}{1+\displaystyle\frac{4m}{r}}\,{\hat p}_k {\hat p}^k+\big(1+\frac{4m}{r}\big)q^2\right),
    \label{(4.4)}
\end{equation}
which is formally the same as the classical expression. Notice, however, that the order of the operators is important.

The charge operator $q$ commutes with the Hamiltonian, and is therefore conserved. The angular momentum operator is
\begin{equation}
        \hat{\vJ}=\hat{\vr}\times\hat{\vp}+4qm\frac{\vr}{r}.
\label{(4.5)}
\end{equation}
Inserting the square of the angular momentum, $J^2$, into the Hamiltonian, we get
\begin{equation}
\hat{H}=\frac{1}{2(1+\displaystyle\frac{4m}{r})}\left(-\frac{1}{r^2}\p_r(r^2\p_r)
+\frac{J^2-(4qm)^2}{r^2}
+\big(1+\frac{4m}{r}\big)^2q^2\right).
\label{(4.6)}
\end{equation}
Substituting here $j(j+1)$ for $J^2$, $s/4m$ for $q$, introducing the new wave function $\psi=r\,\Psi$ and multiplying by $2r(1+4m/r)$, the eigenvalue equation $\hat{H}\Psi=E\Psi$ becomes
\begin{equation}
\left[\frac{\ d^2}{dr^2}-\frac{j(j+1)}{r^2}+\frac{8mE-s^2/2m}{r}-
\Big(\big(\frac{s}{4m}\big)^2-2E\Big)\right]\psi=0.
\label{(4.7)}
\end{equation}
Choosing $\psi\sim r^{j+1}$ as boundary condition near $r=0$, we obtain a
well-defined problem for which (\ref{(4.6)}) is self-adjoint on the Hilbert space
$L^2({\mathbb R}_+, dr)$ over the half-line.
 Continuum solutions of eqn. (\ref{(4.7)}) which are bounded at $r=0$ have the form
\begin{eqnarray}
\psi(r)&=&r^{j+1} e^{\ri kr}F(\ri\lambda+j+1,2j+2,-2\ri kr),
\label{(4.8)}
\\[8pt]
k^2&=&2E-q^2,
\\[8pt]
\lambda&=&-4m\displaystyle\frac{q^2-E}{\sqrt{2E-q^2}}\ ,
\label{(4.9)}
\end{eqnarray}
where $F$  is the confluent hypergeometric function.
 Square-integrable bound states correspond to
 $\lambda$ such that
$\lambda^2=-(k+j+1)^2=-n^2,\  k=0, 1, 2,\dots$,  i.e.,
\begin{equation}
-\frac{4m(q^2-E)}{\sqrt{q^2-2E}}=n,\quad
    n=|s|+1, |s|+2, \dots
    \label{(4.10)}
\end{equation}

For $q^2>2E$ i.e. for $s^2 > 32Em^2$ (which, by eqn. (\ref{(4.4)}), is only possible for negative $m$), one gets the \emph{bound-state energy levels},
\begin{equation}
E_n=\frac{1}{(4m)^2}\sqrt{n^2-s^2}\left(\pm n-\sqrt{n^2-s^2}\right),
\quad n=|s|+1, |s|+2,\dots
\label{(4.11)}
\end{equation}
with degeneracy $n^2-s^2$. Observe that $n^2-s^2=(n+s)(n-s)$ is always an integer, since $n$ and $s$ are simultaneously integers or half-integers. The two signs correspond to the lightly bound states $E>0$ and to the tightly bound states with $E<0$.

    For $E>q^2/2$ one gets scattering states. The quantum cross section can be
calculated \cite{6} by solving the problem in parabolic coordinates. It is identical to
the classical expression (\ref{(3.16)}) with $q/\vert \vec{\bv}\vert $  replaced by $s/4mk$, i.e.,
\begin{equation}
\frac{d\sigma}{d\omega}=\frac{m^2
\left(1+\big({s}/{4mk}\big)^2\right)^2}{\sin^4{\theta}/{2}}\ .
\label{(4.12)}
\end{equation}

Quantizing the Runge-Lenz vector $\vK$ is a hard task. The clue is \cite{8} that $\vK$ is associated with three Killing tensors $K^i_{\mu\nu},\ i = 1, 2, 3$. Remember that a Killing tensor is a symmetric tensor
$K_{\mu\nu}$ on (curved) space, such that $K_{(\mu\nu;\alpha)}=0$ (the semicolon denotes metric-covariant derivative,
$(\, \cdot\,)_;=\nabla_\mu(\, \cdot\,)$).
 To a Killing tensor is associated a conserved quantity which is quadratic in the velocity, namely
$$
K =\frac{1}{2}K_{\mu\nu}\dot{x}^\mu \dot{x}^\nu.
$$
 For example, the metric tensor $g_{\mu\nu}$ itself satisfies these conditions; the corresponding conserved quantity is the energy. Following  Carter \cite{8}, the quantum operator of $K$ is
\begin{equation}
        \widehat{K}=-\frac{1}{2}\nabla_\mu K^{\mu\nu}\nabla_\nu=
        -\frac{1}{2}\nabla^\mu K_{\mu\nu}\nabla^\nu.
\label{(4.13)}
\end{equation}
The classical expression (\ref{(3.8)}) of the Runge-Lenz vector allows us to identify the components of the Killing tensor and use Carter's prescription
to get the quantum operators. A long and complicated calculation yields
\begin{equation}
\hat{\vK}=\frac{1}{2}\big(\hat{\vp}\times\hat{\vJ}-\hat{\vJ}\times\hat{\vp}\big)-4m\frac{\vr}{r}(\hat{H}-q^2).
\label{(4.14)}
\end{equation}
Again, the order of the operators \emph{is} relevant. For notational convenience we drop the `hat' $\hat{\{\,\cdot\,\}}$ from our operators in what follows.

\goodbreak
%%%%%%%%%%%%%%%%%%%%%%%%%%%%%%%%%%%%%%%%%%%%%%%%%
\section{THE SPECTRUM FROM THE DYNAMICAL SYMMETRY}
%%%%%%%%%%%%%%%%%%%%%%%%%%%%%%%%%%%%%%%%%%%%%%%%%

    Using the fundamental commutation relations (\ref{(4.2)}) one verifies that the operators $\vJ$ and $\vK$ satisfy the quantized version of (\ref{(3.17)}),
\begin{equation}
    [J_i,J_k]=\ri\epsilon_{ikn}J^n,
    \qquad
    [J_i,K_k]=\ri\epsilon_{ikn}K^n,
    \qquad
    [K_i,K_k]=\ri(q^2-2H)\epsilon_{ikn}J^n.
\label{(5.1)}
\end{equation}
Analogously to (\ref{(3.18)}),
on the fixed-energy eigenspace $H\Psi=E\Psi$ define the rescaled Runge Lenz operator $\vM$ by
\begin{equation}
\vM\quad=\quad\left.\begin{array}{r}
(q^2-2E)^{-1/2}\vK
\\[6pt]
\vK
\\[6pt]
(2E-q^2)^{-1/2}\vK
\end{array}\right\}
\qquad\hbox{for}\qquad
\left\{\begin{array}{c}
E<q^2/2
 \\[6pt]
E=q^2/2
\\[6pt]
E>q^2/2
\end{array}\right. .
\label{(5.2)}
\end{equation}
$\vM$ and $\vJ$ close, just like classically, to an
${\rm o}(4)$ algebra for $E<q^2/2$,
\begin{equation}
    [J_i,J_k]=\ri\epsilon_{ikn}J^n,
    \qquad
    [J_i,M_k]=\ri\epsilon_{ikn}M^n,
    \qquad
    [M_i,M_k]=\ri\epsilon_{ikn}J^n.
\label{(5.3a)}
\end{equation}
to ${\rm o}(3,1)$ for $E>q^2/2$,
\begin{equation}
    [J_i,J_k]=\ri\epsilon_{ikn}J^n,
    \qquad
    [J_i,M_k]=\ri\epsilon_{ikn}M^n,
    \qquad
    [M_i,M_k]=-\ri\epsilon_{ikn}J^n.
\label{(5.3b)}
\end{equation}
and to  ${\rm o}(3)\oplus_s\IR^3$ in the parabolic case, $E=q^2/2$,
\begin{equation}
    [J_i,J_k]=\ri\epsilon_{ikn}J^n,
    \qquad
    [J_i,M_k]=\ri\epsilon_{ikn}M^n,
    \qquad
    [M_i,M_k]=0.
\label{(5.3c)}
\end{equation}
Following Pauli \cite{9}, this allows us to calculate the energy spectrum.
We have indeed the constraint equations,
\begin{eqnarray}
    \vK\cdot\vJ&=&-(4m)^2q(E-q^2),
\label{(5.4a)}
 \\[8pt]
    \vK^2&=&(2E-q^2)\big(\vJ^2-(4mq)^2+1\big)+4m^2(E-q^2)^2.
\label{(5.4b)}
\end{eqnarray}
Observe that the quantum expressions differ from their classical counterparts,
(\ref{(3.10a)})-(\ref{(3.10b)}), in an important term,
$(h/2\pi)^2=1$ in our units.

Those states $\Psi$ with constant charge, $q=s/4m$, and energy $E<q^2/2$
\begin{equation}
    4mq\,\Psi=s\,\Psi,
    \qquad
    H\Psi=E\Psi
    \label{(5.5)}
\end{equation}
form a representation space for ${\rm o}(4)$. It is more convenient to consider the commuting operators
\begin{equation}
    \vA=\frac{1}{2}\big(\vJ+\vM\big),
    \qquad
    \vB=\frac{1}{2}\big(\vJ-\vM\big),
    \label{(5.6)}
\end{equation}
which generate two independent ${\rm o}(3)$'s,
${\rm o}(4)={\rm o}(3)\oplus{\rm o}(3)$. A common eigenvector
$\Psi$  of the commuting operators $q, H,\vA^2, A_3, \vB^2, B_3$
 satisfies
\begin{equation}
    \vA\,^2\Psi= a(a+1)\Psi,
    \qquad
    A_3\Psi=a_3\Psi,
    \qquad
    \vB\,^2\Psi= b(b+1)\Psi,
    \qquad
    B_3\Psi=b_3\Psi,
\label{(5.7)}
\end{equation}
where $a$ and $b$ are half-integers, and $a_3=-a,-a+1,\dots, a,\;
b_3=-b,-b+1,\dots,b$.

Consider the (so far non-negative real) number
\begin{equation}
    n=-\frac{4m(q^2-E)}{\sqrt{q^2-2E}}.
    \label{(5.8)}
\end{equation}
The operator identities (\ref{(5.4a)}), (\ref{(5.4b)}) imply that
$$
    a(a+1)+b(b+1)=(s^2-1-n),
    \qquad
    a(a+1)-b(b+1)=sn^2.
$$
Then some algebra yields
$$
    2a+1=\pm(n+s),
    \qquad  2b+1=\pm(n-s)   \Rightarrow
    a-b =\pm s,
    \qquad  a+b+1 =\pm n = n,
$$
since $a$ and $b$ are non-negative. Due to the first of these relations, $n$ is integer or half-integer, depending on
$s$ being integer or half-integer. Eqn. (\ref{(5.8)}) is thus identical to eqn. (\ref{(4.10)}), yielding the bound-state spectrum once more.

%%%%%%%%%%%%%%%%%%%%%%%%%%%%%%%%%%%%%%%%%%%%%%%%%%%%%%%%%%
\section{ALGEBRAIC CALCULATION OF THE S MATRIX \cite{10,11}}
%%%%%%%%%%%%%%%%%%%%%%%%%%%%%%%%%%%%%%%%%%%%%%%%%%%%%%%%%%

    The scattering states form rather a representation space for
${\rm o}(3,1)$. This can be used to derive algebraically the S-matrix.
Let us indeed re-write the Runge-Lenz operator $\vK$ as
\begin{equation}
    \vK=\left(1 + 4m/r\right)(\ri\vv-\vJ\times\vv)
    -4m\frac{\vr}{r}(H-q^2),
\label{(6.1)}
\end{equation}
where the velocity operator $\vv$ is defined by
\begin{equation}
\vv=-\ri [\vr, H]= (1+ 4m/r)^{-1} \vp.
\end{equation}

    We consider incoming (respectively outgoing) wave packets, which are sharply peaked
around momentum $\vk$ and have fixed charge
$q = s/4m$ and energy $E=k^2/2+q^2/2$. We argue that they  satisfy the relations
\begin{equation}
(\vJ\cdot\hat{\vk})\Big|\vk\left(\begin{array}{c}
\hbox{\small in}
\\
\hbox{\small out}
\end{array}\right)\Big>
=\mp s \Big|\vk\left(\begin{array}{c}
\hbox{\small in}
\\
\hbox{\small out}
\end{array}\right)\Big>,
\qquad
\hat{\vk}:=\vk/ \vert \vk\vert,
\label{(6.2)}
\end{equation}
\begin{equation}
(\vK\cdot\hat{\vk})\Big|\vk\left(\begin{array}{c}
\hbox{\small in}
\\
\hbox{\small out}
\end{array}\right)\Big>
=(\ri k\pm4m(E-q^2)
\Big|\vk\left(\begin{array}{c}
\hbox{\small in}
\\
\hbox{\small out}
\end{array}\right)\Big>.
\label{(6.3)}
\end{equation}
The states $|\vk(\hbox{\small in})>$ and $|\vk(\hbox{\small out})>$
are solutions of the complete time-dependent Schr\"odinger equation
controlled at $t = \mp\infty$. It is enough to check
the validity of eqn. (\ref{(6.2)}) and (\ref{(6.3)}) at
$t=\mp\infty$, since
$H$ commutes with the operators $\vJ\cdot\hat{\vk}$ and
$\vK\cdot \hat{\vk}$ which are therefore constants of the motion.
Let us apply the left and right hand sides of the operator identity
$\vJ\cdot\frac{\vr}{r}= 4mq$ to the states $|\vk(\hbox{\small in})>$ and
$|\vk(\hbox{\small out})>$ and take into account that
$|\vk(\hbox{\small in})>$
and $|\vk(\hbox{\small out})>$ approach eigenstates of $\frac{\vr}{r}$ with eigenvalues
$\mp \hat{\vk}$ as $t\to \mp \infty$, since $|\vk(\hbox{\small in})>$ and
$|\vk(\hbox{\small out})>$ represent wave-packets incoming from (outgoing to) the direction
 $\hat{\vk}$ when $t\to \mp \infty$. This yields
(\ref{(6.2)}).

    To make (\ref{(6.3)}) plausible, apply rather the operator identity
\begin{equation}
\vK\cdot\hat\vk=\left(1+\frac{4m}{r}\right)\left(\ri \vv\cdot\hat  \vk-
\vJ\cdot(\vv\times\hat{\vk})\right)-4m\frac{\vr}{r}\cdot \hat\vk(H-q^2)
\label{(6.4)}
\end{equation}
to $|\vk(\hbox{\small in})>$ and $|\vk(\hbox{\small out})>$ and notice that,
besides approaching eigenstates of $\vr/r$, they also approach velocity eigenstates with eigenvalue
$\vk$ as $t\to\mp\infty$.
Similarly, the position dependent mass ${\cal M}=1+4m/r$ tends to
unity since the incoming and outgoing wave packets are far away from the monopole's location.
The equations (\ref{(6.2)}) and (\ref{(6.3)}) can also be tested on the explicit
solution of the Schr\"odinger equation, obtained by Gibbons and Manton \cite{6} in parabolic coordinates.

    Having established (\ref{(6.2)}) and (\ref{(6.3)})
 let us now take into account that, for fixed charge and energy, the scattering states span a representation of the dynamical ${\rm o}(3,1)$ algebra. This representation is characterized by the eigenvalues of the Casimir operators, fixed by the constraints (\ref{(5.4a)})-(\ref{(5.4b)}).
    Among those states with fixed energy and charge we have, in addition to those bases which consist of (distorted) plane-wave-like scattering states
    $|\vk(\hbox{\small in})>$ and $|\vk(\hbox{\small out})>$, also the standard angular momentum basis (spherical waves) $|j, j_3 > \ ,\
    (j = |s|,|s|+1,\dots )$,
\begin{equation}
\vJ\,^2|j,j_3>=j(j+1)|j,j_3>\,,
\qquad
J_3|j,j_3>=j_3|j,j_3>\,.
\label{(6.5)}
\end{equation}
The angular momentum basis is easy to handle. We would obtain the matrix elements
\hfill\break
$<\vec{l}(\hbox{\small out})|\vk(\hbox{\small in})>$ if we could expand
$|\vk(\hbox{\small in})>$ and $|\vk(\hbox{\small out})>$ in the angular momentum basis.
The desired expansions can be algebraically derived, since one has the ${\rm o}(3,1)$ algebra and all the basis vectors
$|\vk(\hbox{\small in})>$\,,\ $|\vk(\hbox{\small out})>$\ ,
 and $|j, j_3>$ are eigenvectors of the appropriate components of the generators. The details are practically identical to those presented by Zwanziger \cite{10}. This leads to the final expression
\begin{equation}
S\big(\vec{l},\vk\big)=\big<\vec{l}(\hbox{\small out})|\vk(\hbox{\small in})\big>=
\sum_{j\geq \vert s\vert }(2j+1) \left[\frac{(j-\ri\lambda)!}{(j+\ri\lambda)!}\right]
{\cal D}^{\ j\ }_{(-s,s)}\big(R^{-1}_{\vec{l}}R_\vk\big),
\label{(6.6)}
\end{equation}
with $\lambda$ given in (\ref{(4.9)}). The ${\cal D}$'s are
the rotation matrices and $R_{\vk}$ is a rotation which brings the
$z$ direction into the direction of $\vk$.
Eq.~(\ref{(6.6)}) is consistent with Eq.~(\ref{(4.12)}) for the
cross-section, and its poles yield the bound-state spectrum (\ref{(4.11)}).

    It should be emphasized that the derivation of the S-matrix given here above crucially
    depends on the expression (\ref{(6.1)}) of the Runge-Lenz vector $\vK$ in terms of the 3 dimensional variables and on the constraint equations
(\ref{(5.4a)})-(\ref{(5.4b)}), which fix the actual representation of ${\rm o}(3,1)$. These relations are in turn straightforward consequences of Carter's covariant expressions.

%%%%%%%%%%%%%%%%%%%%%%%%%%%%%%
\section{EXTENSION INTO ${\rm o}(4,2)$}
%%%%%%%%%%%%%%%%%%%%%%%%%%%%%%

Now we extend \cite{12,13} the ${\rm o}(4)$ symmetry into the conformal
algebra ${\rm o}(4,2)$. Let us re-arrange the time independent
Schr\"odinger equation as
\begin{equation}
\left( (1+4m/r)^{-1} \vp\,{\strut}^2+\big(1+4m/r\big)q^2-2E\right)\Psi=0.
\label{(7.1)}
\end{equation}
When multiplied by $r(1+4m/r)=r+4m$ from the left, the operator on
the l.h.s. becomes, after rearrangement,
\begin{equation}
r\,\vp\,{\strut}^2+r\big(q^2-2E\big)+\frac{(4mq)^2}{r}=8m(E-q^2).
\label{(7.2)}
\end{equation}

Let us work first with bound motions, $q^2/2>E$,
and introduce the variables
\begin{equation}
\vR=\vr\,(q^2-2E)^{1/2}
\qquad\hbox{and}\qquad
\vP=\vp\,(q^2-2E)^{-1/2}\ .
\label{(7.3)}
\end{equation}
The new operators $\vR$ and $\vP$ satisfy the commutation relations
\begin{equation}
[P_j,P_k]=-\ri(4mq)\epsilon_{jkn}\frac{R^n}{R^3},
\qquad
[R^j,R^k]=0,
\qquad
[R^j,P_k]=\ri\delta^j_{\ k}.
\label{(7.4)}
\end{equation}
In terms of $\vR$ and $\vP$, eqn. (\ref{(7.1)}) becomes
\begin{equation}
R(\vP\,^2+1)+\frac{(4qm)^2}{R}=-8m\frac{E-q^2}{\sqrt{q^2-2E}}\ .
\label{(7.5)}
\end{equation}
On the left hand side of (\ref{(7.5)}) we recognize
$\Gamma_0$, the generator of an ${\rm o}(2,1)$ algebra.
Indeed, as a consequence of the commutation relations
(\ref{(7.5)}), the operators
\begin{eqnarray}
\Gamma_0&=&\frac{1}{2}\big(R\vP^2+R+\frac{(4qm)^2}{R}\big),
\label{(7.6a)}
\\[6pt]
\Gamma_4&=&\frac{1}{2}\big(R\vP^2-R+\frac{(4qm)^2}{R}\big),
\label{(7.6b)}
\\[6pt]
D&=&\vR\cdot\vP-\ri
\label{(7.6c)}
\end{eqnarray}
satisfy the ${\rm o}(2,1)$ relations
$$
[\Gamma_0,\Gamma_4]=\ri D,
\qquad [\Gamma_4,D]=-\ri \Gamma_0,
\qquad
[D,\Gamma_0] = \ri \Gamma_4.
$$
 The energy spectrum (but not the degeneracy) is recovered from this at once: the ${\rm o}(2,1)$ generator $\Gamma_0$ has eigenvalues
 $n = |s|+1, |s|+2,\dots$ Equating the r.h.s. of eqn. (\ref{(7.5)}) with $2n$, we get once more the crucial relation
 (\ref{(4.10)}). Let us complete (\ref{(7.6a)})-(\ref{(7.6c)})
with $12$ more operators, namely with
\begin{eqnarray}
\vV &=&R\vP
\label{(7.6d)}
\\[8pt]
\vJ &=& \vR \times \vP + 4qm \frac{\vR}{R},
\label{(7.6J)}
\\[8pt]
\vM&=&\frac{1}{2}\vR(\vP)^2-\vP(\vR\cdot\vP)-\frac{\vR}{2}-4qm\frac{\vJ}{R}+
(4qm)^2\frac{\vR}{2R^2},
\label{(7.6f)}
\\[8pt]
\vU&=&\frac{1}{2}\vR(\vP)^2-\vP(\vR\cdot\vP)+\frac{\vR}{2}-4qm\frac{\vJ}{R}+
(4qm)^2\frac{\vR}{2R^2}.
\label{(7.6g)}
\end{eqnarray}
The commutation relations (\ref{(7.4)}) imply that these operators extend the ${\rm o}(2,1)$ algebra (\ref{(7.6a)})-(\ref{(7.6c)}) into an ${\rm o}(4,2)$ conformal algebra. Those operators commuting with $\Gamma_0$ form an
${\rm o}(2)\oplus{\rm o}(4)$, generated by $\Gamma_0$ itself and by $\vJ$ and
$\vM$. Expressed in terms of $\vr$ and $\vp$, we see that
$\vJ$ is just the angular momentum operator. On the other hand,
$\vM$ is also written as
\begin{equation}
\vM=\frac{1}{2}\big(\vP\times\vJ -\vJ\times\vP\big) + \frac{\vR}{R}\,\Gamma_0.\label{(7.7)}
\end{equation}
Using the relation (\ref{(7.5)}) this becomes
\begin{equation}
\vM=\frac{1}{2}\left(\vP\times\vJ-\vJ\times\vP\right)-4m\frac{\vR}{R}\,
\frac{E-q^2}{\sqrt{q^2-2E}}\ .
\label{(7.8)}
\end{equation}
Substituting here $\vr$ and $\vp$,
$\vM$ reduces to (the restriction onto $H\Psi=E\Psi$ states) of the rescaled Runge-Lenz vector in (\ref{(5.2)}), as anticipated by our notation.
We conclude that the ${\rm o}(4,2)$ algebra (\ref{(7.6a)})-(\ref{(7.6g)})
 extends the original dynamical ${\rm o}(4)$ symmetry.

The scattering case $E>q^2/2$ is treated exactly the same way. Defining
\begin{equation}
\vR=\vr\, \sqrt{2E-q^2}\, ,
\label{(7.9)}
\end{equation}
one discovers, after suitable rearrangement, the non-compact operator
$\Gamma_4$ with continuous spectrum. The extension to ${\rm o}(4,2)$ proceeds along the same lines as before. Those operators commuting with
$\Gamma_4$ form ${\rm o}(2)\oplus{\rm o}(3,1)$, generated by $\Gamma_4, \vJ$, and
$\vU$, this latter being now identified with the rescaled Runge-Lenz vector
$\vM$.
    Note that a similar procedure works for a particle in a self-dual monopole field \cite{14}.

\kikezd{Note added in 2009.} This review  has been presented at the
1988 Schloss Hofen Meeting on {\it Symmetries in Science III}, and
published in the Proceedings: {\it Symmetries in Science III},
Gruber B and Iachello F (eds),  Plenum,  New York, pp.~399-417,
(1989). Neither the original text nor the list of references have
been updated. We remark that the  somewhat heuristic treatment
 in Section 7 can be made rigorous along the lines
 followed in the paper~: B. Cordani, L. Feh\'er, and P. A. Horv\'athy~:
``Kepler-type dynamical symmetries of long-range monopole
interactions.'' {\sl Journ. Math. Phys.} {\bf 31}, 202 (1990), where
the classical aspects were further investigated.

\kikezd{ACKNOWLEDGEMENTS.}
    Some of the results presented here were obtained in collaboration with B. Cordani, to whom we express our indebtedness. We also thank P. Forg\'acs, G. Gibbons, Z. Horv\'ath, L. O'Raifeartaigh and L. Palla for discussions.
We are indebted to M. Perry for granting us his permission to follow  Ref. \cite{2} in Sections 1 and 2.

\end{document}